# Biphoton interference with a quantum dot entangled light source

R. M. Stevenson[1*], A. J. Hudson[1,2], R. J. Young[1], P. Atkinson[2], K. Cooper[2],
D. A. Ritchie[2], & A. J. Shields[1]

[1]*Toshiba Research Europe Limited, 260 Cambridge Science Park,
Cambridge CB4 0WE, UK*

[2]*Cavendish Laboratory, University of Cambridge, Madingley Road,
Cambridge CB3 0HE, UK*

*Corresponding author: mark.stevenson@crl.toshiba.co.uk

**Abstract:** We demonstrate optical interferometry beyond the limits imposed by the photon wavelength using 'triggered' entangled photon pairs from a semiconductor quantum dot. Interference fringes of the entangled biphoton state reveals a periodicity half of that obtained with the single photon, and much less than that of the pump laser. High fringe visibility indicates that biphoton interference is less sensitive to decoherence than interference of two sequential single photons. The results suggest that quantum interferometry may be possible using a semiconductor LED-like device.

**1. Introduction**

Two-photon interferometry is a powerful technique that typically employs coincident detection of pairs of photons to probe key characteristics of non-classical light. Such properties include the coherence length and the de Broglie wavelength, which determines the far-field imaging resolution [1]. The nature of two-photon interference can be remarkably different from that of the constituent single photons. For example, higher frequency interference fringes have been observed using entangled photons generated by parametric down conversion [2,3,4], or by post-selected measurements using classical light [5], which is the basis of quantum imaging applications such as quantum lithography [6,2], and low-cell-damage biomedical microscopy [7]. Triggered entangled photon pair generation by quantum dots has recently been demonstrated [8,9], and here we study biphoton interference of the emission from such a device. A primary motivation is to observe fringes with finer detail than is possible with the pump laser, in contrast to pairs generated by parametric down conversion.

Single and two-photon interference has been measured previously for sequentially emitted single photons from individual quantum dots [10,11]. Interference of single photons has revealed that the photon coherence time is much shorter than the radiative lifetime, which limits the interference visibility between two sequentially emitted photons at the same wavelength. Thus in order to generate entangled photon pairs via two-photon interference [12], photons are made more indistinguishable by the suppression of single photon decoherence, usually by resonant excitation. In contrast, biphoton interferometry is sensitive to the coherence between superposed components of the entangled photon-pair state, and the individual photons need not be indistinguishable nor have the same wavelength. The corresponding measurement or setting of phase with the interferometer thus accesses the phase between the orthogonally polarized components of the entangled biphoton. We present below interference experiments that suggests two-photon decoherence does not limit the entanglement of a photon pair, in contrast to the effect of single-photon decoherence on entanglement produced by interference of sequentially emitted photons. In addition, we determine and also control the phase offset of the entangled state.

## 2. Methods

A quantum dot can emit a pair of polarisation entangled photons by the radiative decay of the biexciton (XX) state providing the intermediate exciton (X) level is degenerate, as shown schematically in Fig. 1(a). Typically, structural asymmetries such as shape and strain lead to polarisation dependent splitting of the X state, and emission of only classically polarisation correlated photons [13-15]. However, two schemes to eliminate the polarisation splitting by control of growth [16] or application of an in-plane magnetic field [17], have enabled the emission of entangled photons. Further promising schemes are emerging, such as annealing a selected dot [18,19], or by application of external strain [20] or electric field [21,22]. An alternative strategy to generate entangled photons with quantum dots employs energetic post-selection of the emitted photons [23].

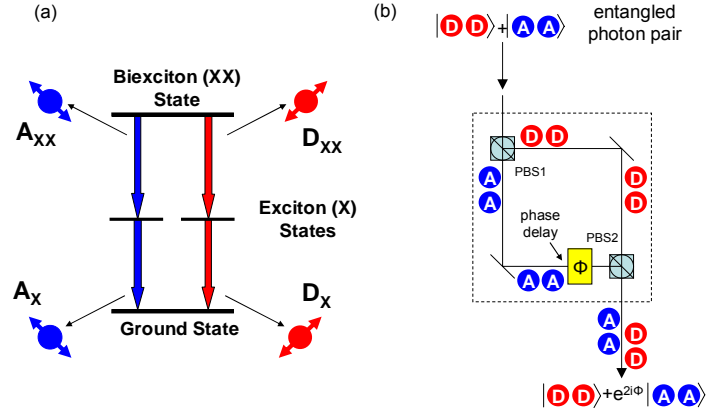

Fig. 1. (a). Energy level diagram of radiative decay of biexciton state in a quantum dot. (b) Schematic of a biphoton interferometer.

For the measurements presented in this work, growth control with molecular beam epitaxy was used to fabricate a dot with polarisation splitting 0.0±0.6µeV, which emits entangled photons. A single layer of InAs quantum dots was formed at the centre of a λ GaAs cavity, with GaAs/AlAs distributed Bragg reflectors below (14 repeats) and above (2 repeats). The thickness of the InAs layer was optimised to achieve a dot density of ~1µm$^{-2}$, with lower background light emission compared to earlier designs [9]. Apertures ~ 2µm in diameter were fabricated in a metal shadow mask on the sample surface in order to isolate single quantum dots. Photon pairs emitted by the source have 76±4% fidelity with the Bell state $(|H_{XX}H_X\rangle+|V_{XX}V_X\rangle)/\sqrt{2}$ and also with the equivalent Bell state $(|D_{XX}D_X\rangle+|A_{XX}A_X\rangle)/\sqrt{2}$, where $H$, $V$, $D$, and $A$ represent horizontal, vertical, diagonal, and anti-diagonal linear polarisation, of the XX and X photons. This is the highest fidelity for this type of source reported to date. Biphoton interference was measured using an interferometer similar to that shown in Fig. 1(b). The phase control required to measure an interferogram is supplied by a polarisation dependent phase delay. A combination of appropriately configured polarising beamsplitters PBS1 and PBS2 force $D$ and $A$ polarised photons to take separate paths through the interferometer. Both $A$ polarised photons are delayed by phase $\Phi$, so that output of the interferometer is $(|D_{XX}D_X\rangle+e^{2i\Phi}|A_{XX}A_X\rangle)/\sqrt{2}$. Interference of the two-photon amplitudes of $|D_{XX}D_X\rangle$ and $|A_{XX}A_X\rangle$ is achieved by measuring the projection onto the two-photon states $|V_{XX}V_X\rangle$ or $|V_{XX}H_X\rangle$. The biphoton intensity variation with $\Phi$ results in an interferogram, with an expected period $\pi$, corresponding to a de Broglie wavelength of $\lambda/2$, where $\lambda$ is the average wavelength of the single biexciton and exciton photons $|A_{XX}\rangle$ and $|A_X\rangle$. For the single photon input state $|V\rangle=(|D\rangle+|A\rangle)/\sqrt{2}$, the output from the interferometer is

$(|D\rangle+e^{i\Phi}|A\rangle)/\sqrt{2}$. Measuring the intensity by the projection onto the state $|V\rangle$ would yield an interferogram with period $2\pi$, corresponding to a de Broglie wavelength $\lambda$.

The required polarisation dependent phase delay was realised by a liquid crystal with voltage dependent birefringence. Photons polarised along the diagonally oriented slow axis of the crystal are delayed relative to those polarised along the anti-diagonally oriented fast axis. The collinear nature of the interferometer [5] provides exceptional stability compared to a Mach-Zehnder arrangement where photons are spatially separated according to their polarisation. The error in the polarization dependent phase delay was determined to be $0.03\lambda$.

Interference between photons travelling on the fast and slow paths is achieved by detection in rectilinear polarisation. A non-polarising beam splitter directs the output of the interferometer into two spectrometers, one set to filter light resonant with biexciton emission, and one resonant with exciton emission. A polarising beam splitter directs the output of the exciton spectrometer to a pair of silicon avalanche photo-diodes (APDs). A linear polariser selects the polarisation of filtered biexciton light, before detection by another APD. Each APD can measure polarised single photon intensities for the exciton or biexciton.

## 3. Interference measurements

We first measure single photon interference fringes from light emitted by a quantum dot. The dot is optically excited non-resonantly at a temperature of 10K, at 632nm with a frequency of 80MHz. A linear polariser is inserted before the interferometer to select only vertically polarised photons. The intensity of the single exciton photon state $|V_X\rangle$ is measured as a function of the phase delay, and normalised to the maximum. The results are shown in Fig. 2 as black points. Clear interference fringes are seen, and the intensity varies as a function of the phase delay in agreement with the fit to expected sinusoidal behaviour, shown by the solid line. The period of the oscillations is determined to be 877±35nm (0.99±0.03)$\lambda$, approximately equal to the wavelength of the quantum dot emission of ~885nm, as expected.

The linear polariser set before the interferometer was removed, so that entangled photon pairs emitted by the quantum dot could be analysed. The normalised biphoton intensity is equal to $(g_{VV}+g_{HH})/(g_{VV}+g_{VH}+g_{HV}+g_{HH})$, where the denominator and numerator and are proportional to the biphoton generation and detection rates respectively, and $g$ represents the second order correlation for coincident detection of two photons with polarisation denoted by subscripts. For an unpolarised source, such as this dot, this is approximated to $g_{VV}/(g_{VV}+g_{HV})$. The second order correlation functions for the $|V_{XX}V_X\rangle$ and $|H_{XX}V_X\rangle$ two-photon detection bases were measured simultaneously as described elsewhere [9].

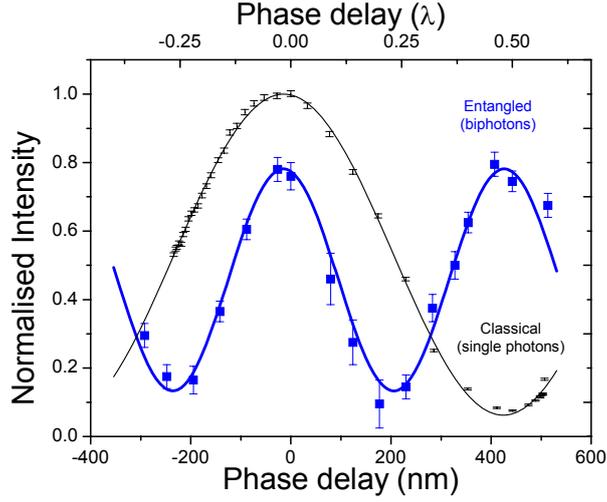

Fig. 2. Normalised intensity of classical single photons (black) and normalised biphoton intensity of entangled photon pairs (blue) as a function of the phase delay.

The measured normalised biphoton intensity, indicated by blue points in Fig. 2, shows strong interference fringes. The difference in the period of oscillations compared to the classical single photon case is very striking. The fringes fit well to the predicted sinusoidal behaviour shown as a solid line (coefficient of determination $r^2$=94.4%), from which we determine the period of the oscillations to be 442±36nm (0.50±0.03)$\lambda$. The period is equivalent to the de Broglie wavelength of the biphoton, which is in excellent agreement with the two-fold reduction from 885nm to 443nm expected for an entangled photon pair source.

The shorter de Broglie wavelength for the entangled state compared to the single photon implies that up to a two-fold enhancement of the imaging resolution is possible using biphoton detection. The visibility of the measured biphoton interferogram compared to that for single photon detection demonstrates enhanced phase resolution of the interferometer with a biphoton source [7]. Note that the exciton and biexciton photons need not have the same wavelength in order to observe biphoton interference [24], as it is the total energy of the two-photon state that defines the de Broglie wavelength. This is the usual case for quantum dots.

From the fit to sinusoidal behaviour we also directly determine the phase difference between the $|D_{XX}D_X\rangle$ and $|A_{XX}A_X\rangle$ components of the entangled state emitted by the source to be 0.02±0.03λ. To our knowledge, this represents the most direct measurement of phase offset reported for entangled photons generated by a quantum dot, and shows that the diagonally polarised two-photon amplitudes are in-phase within error.

The fact that we can control the phase offset also means that the entangled state can be manipulated into another form. For example, a λ/4 delay transforms the state $(|H_{XX}H_X\rangle+|V_{XX}V_X\rangle)/\sqrt{2}$ into $(|H_{XX}V_X\rangle+|V_{XX}H_X\rangle)/\sqrt{2}$, and the photons are then polarisation anti-correlated in the rectilinear basis.

The above results show clear differences between the interference fringes measured for single photons and entangled photon pairs. Next we compare the results for the entangled state, with other photon pair states that are classical in nature. First, we measure interference fringes for the classical state $|V_{XX}V_X\rangle$, which is itself a component of the maximally entangled state $(|H_{XX}H_X\rangle+|V_{XX}V_X\rangle)/\sqrt{2}$. $|V_{XX}V_X\rangle$ is a pure classical state, and thus can be written as a product of its component exciton and biexciton single photon states as $|V_{XX}\rangle|V_X\rangle$. We measure the exciton and biexciton intensities independently and multiply to determine the interferogram for the state $|V_{XX}V_X\rangle$, selected by passing the dot emission through a linear polariser. The corresponding interference fringes, shown in Fig. 3 by black discs, deviate significantly from the sinusoidal variation observed in Fig. 2, and instead agree well with fits proportional to $(1+\cos(\Phi))^2$ shown by the solid black line, as expected for these particular states. As expected, the measurements agree well with direct detection of the biphoton intensity [25] for $|V_{XX}V_X\rangle$, shown by black squares in Fig. 3.

Secondly, we consider whether a classical mixture, rather than superposition, of $|H_{XX}H_X\rangle$ and $|V_{XX}V_X\rangle$ can reproduce the observed behaviour of the entangled state. In ordinary quantum dots, the lack of a degenerate exciton state results in only classical polarisation correlated photon pair emission. In an idealised case, the photon pairs emit into the mixed state consisting of equal parts of $|H_{XX}H_X\rangle$ and $|V_{XX}V_X\rangle$. We suppress the emission of entangled photons in the quantum dot under investigation, by the application of an in-plane magnetic field [8,17], which increases the exciton splitting S from 0.0±0.6μeV at 0T to 24.8±0.6μeV at 4T. The resulting emission is expected to be highly classical, as reported previously [8]. We measure the biphoton

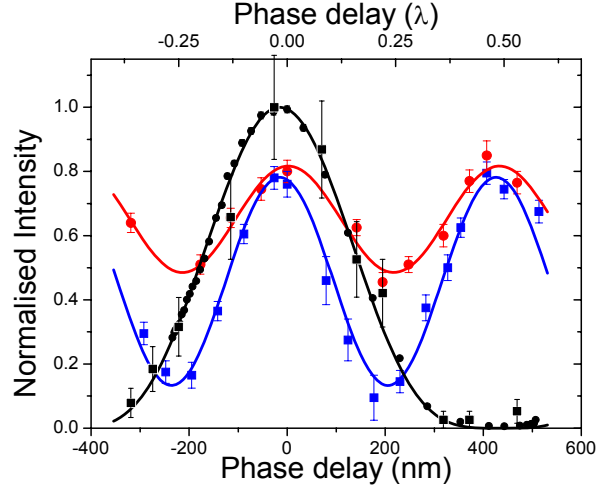

Fig. 3. Biphoton inteferogram for the pure classical state $|V_{XX}V_X\rangle$ (black), and a mixed classical state from an unentangled dot (red). The mixed state is polarisation correlated, with strong components $|V_{XX}V_X\rangle$ and $|H_{XX}H_X\rangle$. The interferogram for the entangled state (blue) shows half the period, and double the visibility compared to the pure and mixed classical states respectively.

intensity of the emission as a function of phase delay, and compare with that measured at 0T in Fig. 3. The visibility of the interference fringes for the classical state shown in red is very different to that shown by the entangled state shown in blue. In fact, the amplitude of the fringes is only 0.34±0.03, within error of half that measured at 0T of 0.65±0.04. In addition, the intensity corresponding to the classical state never dips below 0.5, indicating the absence of destructive interference.

The form of the interference can be understood by considering incoherent mixing of the interference patterns for the component states $|V_{XX}V_X\rangle$ (shown in black), and $|H_{XX}H_X\rangle$, which is similar shape, but offset in phase by $\pi$ (not shown). This would result in a sinusoidal ripple in the intensity with amplitude 0.5 and period $\lambda/2$, on top of a constant background of 0.5. A sinusoidal fit agrees well with the measurements, though the visibility observed is reduced due to the presence of background light and exciton spin scattering [13,14]. Equivalent degradation of the entangled state would yield a ratio between the visibilities of the interference at 0T and 4T of 2, consistent with the factor 1.91±0.21 we observe. The low visibility of the fringes for a mixed classical state also means that no resolution enhancement would be achieved with classical two-photon imaging, despite the observation of fringes with period $\lambda/2$. We note for a period of $\lambda/2$, biphoton interference amplitudes exceeding 50% can only be explained non-classically [26].

Uncorrelated light originates from other layers of the sample, and also from scattering between the intermediate exciton spin states [13,14], and reduces the amplitude of the interference maxima for entangled or mixed classical light, and the interference minima for entangled light. From the interference of the mixed classical state, we determine that uncorrelated light contributes towards (37±7)% of the biphoton intensity. The corresponding maximum possible fidelity and interference amplitudes for the entangled state are 0.73±0.05 and 0.63±0.07 respectively, in close agreement with measured values.

Biphoton decoherence is the decay of coherence between the polarised superposed components of the entangled biphoton. Though we do not directly measure the biphoton coherence time, the agreement between the expected maximum, and measured values for both fidelity and interference amplitude suggests that biphoton decoherence does not degrade the entangled state. In contrast, the single photon coherence time limits the visibility of two-photon interference using two sequential photons from a quantum dot, and necessitates the use of resonant excitation [10,11].

## 4. Conclusions

Strong biphoton interference has been observed for triggered entangled photon pairs from a quantum dot. Decoherence has little effect on the visibility, in contrast to two-photon interference of successively emitted single photons from a dot. The phase difference between the emitted polarisation components of the biphoton is found to be zero. By manipulating this relative phase, we can create different entangled biphoton states. Ultimately it may be possible to implement the source using a simple LED-like design [27], allowing the practical realization of quantum enhanced interferometry.

**Acknowledgements**

We would like to thank the EU projects QAP and SANDiE, the EPSRC, and the QIP IRC for funding.

**References and links**